\documentclass[aps,prl,showpacs,twocolumn]{revtex4-2}
\usepackage{graphicx}
\usepackage{subfig}
\usepackage{tabularx}
\usepackage{bm}
\usepackage{amsmath}
\usepackage{esvect}

\providecommand{\abs}[1]{\left\lvert#1\right\rvert}

\providecommand{\bra}[1]{\langle #1 \rvert}
\providecommand{\ket}[1]{\lvert #1 \rangle}

\providecommand{\be}{\begin{equation}}
\providecommand{\ee}{\end{equation}}
\providecommand{\ba}{\begin{eqnarray}}
\providecommand{\ea}{\end{eqnarray}}

\usepackage{mathbbol}
\def\N{\mathbb{N}}

\def\C{\mathds{C}}

\usepackage{tikz}
\usepackage{amsfonts,amssymb}
\usepackage{dsfont}
\usepackage{physics}
\usepackage{tocvsec2}

\usepackage{graphicx}
\usepackage{subfig}
\usepackage{tabularx}
\usepackage{bm}
\usepackage{amsmath}
\usepackage{esvect}

\usepackage{siunitx}
\usepackage{hyperref}
\usepackage[capitalize]{cleveref}

\providecommand{\abs}[1]{\left\lvert#1\right\rvert}

\providecommand{\bra}[1]{\langle #1 \rvert}
\providecommand{\ket}[1]{\lvert #1 \rangle}

\providecommand{\be}{\begin{equation}}
\providecommand{\ee}{\end{equation}}
\providecommand{\ba}{\begin{eqnarray}}
\providecommand{\ea}{\end{eqnarray}}
\usepackage{amsfonts,amssymb}
\usepackage{dsfont}
\usepackage{physics}
\usepackage{tocvsec2}

\hypersetup{colorlinks=true,linkcolor=blue,citecolor = blue, urlcolor=black}
\newcommand{\beq}{\begin{equation}}
\newcommand{\eeq}{\end{equation}}

\usepackage{appendix}

\begin{document}

\title{Superselection rules and bosonic quantum computational resources}
 
\author{ Eloi Descamps$^{1}$}
\author{ Nicolas Fabre$^{2}$}
\author{ Astghik Saharyan$^{1}$}
\author{ Arne Keller$^{1,3}$}
\author{Pérola Milman$^{1}$ }
\email{corresponding author: perola.milman@u-paris.fr}

\affiliation{$^{1}$Laboratoire Mat\'eriaux et Ph\'enom\`enes Quantiques, Universit\'e Paris Cit\'e, CNRS UMR 7162, 75013, Paris, France}
\affiliation{$^{2}$ Télécom Paris-LTCI, Institut Polytechnique de Paris, 19 Place Marguerite Perey, 91120 Palaiseau, France}
\affiliation{$^{3}$Departement de Physique, Université Paris-Saclay, 91405 Orsay Cedex, France}

\begin{abstract}
We present a method to systematically identify and classify quantum optical non-classical states as classical/non-classical based on the resources they create on a bosonic quantum computer. This is achieved by converting arbitrary bosonic states into multiple modes, each occupied by a single photon, thereby defining qubits of a bosonic quantum computer. Starting from a bosonic classical-like state in a representation that explicitly respects particle number super-selection rules, we apply universal gates to create arbitrary superpositions of states with the same total particle number. The non-classicality of the corresponding states can then be associated to the operations they induce in the quantum computer. We also provide a correspondence between the adopted representation and the more conventional one in quantum optics, where superpositions of Fock states describe quantum optical states, and we identify how multi-mode states can lead to quantum advantage. Our work contributes to establish a seamless transition from continuous to discrete properties of quantum optics while laying the grounds for a description of non-classicality and quantum computational advantage that is applicable to spin systems as well.

\end{abstract}
\vskip2pc 
 
\maketitle

Quantum theory leads to classically counter-intuitive phenomena not only in physics, but also in computer science \cite{nielsen00}. In particular, certain quantum algorithms have been shown to significantly outperform their classical counterparts \cite{v009a004, Shor}. This leads to the definition of {\it quantum advantage} - the inability to efficiently simulate a given protocol using classical computational resources -, which serves as a well-suited criterion for non-classicality from a computer science perspective.  Hoever, from a physical perspective, non-classicality is not so clearly defined, since it is often associated to different properties, such as entanglement and non-locality \cite{Bell, CHSH} and, in quantum optics, to the field's statistical properties \cite{PhysRevLett.39.691,PhysRevLett.51.384,PhysRevLett.55.2409, DeBievre1}. Several non-classicality criteria exist, each depending on different aspects of a quantum state \cite{PhysRevLett.122.080402, NC, article,PhysRevA.72.043808, PhysRevLett.123.043601}, such as its phase space representation \cite{AnatoleKenfack_2004, PhysRevLett.89.283601,Stellar, DeBievre2PhaseSpace,PhysRevResearch.5.L032006} or its metrological properties \cite{Hillery,Nonclassical2,Nonclassical1, Bouchard:17}, sometimes yielding seemingly paradoxical classifications \cite{Goldberg1}. 

The development of quantum information theory and the possibility to encode quantum information in bosonic fields \cite{Braunstein, KLM, Lloyd:1996aai} have resulted in significant progress in understanding and combining the physical and computational aspects of non-classical phenomena. Does the classical/non-classical frontier coincide for both domains? A crucial result towards answering to this question was the identification of a set of universal gates for position and momentum-like variables (as the quadratures of the electromagnetic field) \cite{Braunstein}. Using this set of gates, unitaries efficiently connecting (in polynomial runtime) any two quantum optical states with arbitrary precision can be constructed. However, it was shown that the restriction to Gaussian states (as coherent or squeezed states, that have a non-negative Wigner function \cite{PhysRev.40.749}) manipulated by Gaussian operations can be efficiently classically simulated \cite{PhysRevLett.109.230503}. Consequently, a quantum computer built only with optical Gaussian components cannot result into quantum advantage. These results establish non-Gaussianity as a necessary condition for quantum optical systems to function as ``hardware" that enables quantum advantage. Nevertheless, non-Gaussianity alone - whether in states or operations - is not sufficient to ensure the impossibility of classical simulation \cite{Calcluth2022efficientsimulation, calcluth2023sufficient, PhysRevResearch.3.033018}. Consequently,  it does not appear as a clear common physical and computational criterium separating classical from non-classical phenomena. 

Sufficient conditions for universality and potential quantum advantage for different bosonic encodings were established from an informational resource theoretical perspective \cite{calcluth2023sufficient}. In addition, the classical computational resources required for the simulation of a quantum circuit with different input states \cite{Simulation, PhysRevA.98.062314} were quantified in \cite{PhysRevLett.130.090602} in terms of a non-Gaussianity measure, the stellar rank \cite{Stellar}, leading to the conclusion that it is rather non-Gaussian entanglement \cite{MattiaX} that is a necessary ingredient for quantum computational advantage. Despite these notable advancements, the questions remain: Is it possible to consistently define and classify general bosonic states in terms of their quantum computational power, {\it i.e.}, in terms of the way these states are translated into operations on a quantum computer? Is it possible to establish a criterion of non-classicality for bosonic states that is physically sound and computationally meaningful, thereby providing a coherent framework that can be applied to {\it arbitrary} states? In this case, how can different bosonic encodings be interpreted and compared within the provided framework?

In this Letter, we address these questions by introducing a systematic method to map general bosonic quantum states onto a digital (qubit-based) quantum computer. This approach enables the interpretation of any bosonic quantum state from a unifying informational perspective. Specifically, we demonstrate how the application of universal gates on an initially classical state translates into qubit operations within a bosonic quantum computer (BQC). In a $N$ qubit BQC, qubits are encoded in modes: modes associated with external degrees of freedom, as the propagation direction, label each one of the $N$ qubits, and those associated with internal ones, as two orthogonal polarizations, represent the qubit states. Each one of the $N$ propagation modes is occupied by one photon, split into two polarizations, as we detail in the following. The concepts of universality and quantum advantage are well established for BQCs \cite{KLM}. We will then identify classical-like quantum optical states and operations to a state space of the BQC that can be efficiently classically simulated (in particular using classical optics). In contrast, non-classical states will be transformed into operations that promote the BQC to universality and render it not efficiently classically simulable.

Central to our approach is the adoption of a particle number superselection rule-compliant (SSRC) description of quantum optical states \cite{PhysRevA.68.042329, PhysRevA.55.3195, PhysRevA.58.4244,PhysRevA.58.4247,doi:10.1142/S0219749906001591,Sanders_2012,PhysRev.155.1428}, which forbids the existence of states that involve superpositions of different total particle numbers. This framework, which we now summarize, has been the subject of a fruitful debate \cite{doi:10.1142/S0219749906001591}. In the SSRC representation, bosonic states are described as two-mode states, with one mode serving as an internalized phase reference \cite{RevModPhys.79.555}. A general SSRC pure state can be expressed as 
\be\label{Psi}
\ket{\Psi}= \sum_{n=0}^{N} c_n\ket{n}_{\bf A}\ket{N-n}_{{\bf R}},
\ee
where ${\bf A} ({\bf R})$ are orthogonal modes (say, spatial), $c_n$ are arbitrary complex coefficients, $\ket{N}_{{\bf A} ({\bf R})}= \left (\hat a_{{\bf A} ({\bf R})}^{\dagger} \right )^N/\sqrt{N!}\ket{\emptyset}$, and $\ket{\emptyset}$ is the vacuum state. The phase reference is said to be {\it internalized}, or {\it quantized}, since it is associated to a Hilbert space ${\cal H}_{\bf R}$.

The SSRC representation describes exactly the same physical phenomena as the standard representation where pure bosonic states consist of coherent superpositions of Fock states.  We will from now on refer to the representation where SSRs are  not explicitly observed as the continuous variables (CV) one.  Moving from the SSRC representation \eqref{Psi} to the CV one is formally possible by defining a mapping ${\cal M}$ such that $\ket{n}_{\bf A}\ket{N-n}_{{\bf R}} \mapsto \ket{n}_{\bf G}\ket{N}_{{\bf K}}$, (restricted to $n \leq N$) where the Hilbert space ${\cal H}_{\bf G}\otimes {\cal H}_{\bf K}$ is isomorphous to ${\cal H}_{\bf A}\otimes {\cal H}_{\bf R}$, and  ${\bf G}$ and ${\bf K}$ are orthogonal modes \cite{RevModPhys.79.555}. Hence, applying ${\cal M}$ to \eqref{Psi},  $\ket{\Psi} \mapsto \ket{\Psi'}$, where $\ket{\Psi'}=\sum_{n=0}^{N} c_n\ket{n}_{{\bf G}}\ket{N}_{\bf K}$, which is a separable state. We then recover the ``usual" CV description by tracing out state $\ket{N}_{\bf K}$. In this case, the phase reference is said to be {\it externalized}, or turned into classical. 
To physically illustrate how the classical treatment of a phase reference leads to the superposition of Fock states, we can refer to the generation of a coherent state \cite{PhysRev.131.2766} $\ket{\alpha}=e^{-|\alpha|^2/2}\sum_{n=0}^{\infty} \alpha^n/\sqrt{n!}\ket{n}$ by a classical oscillating current: if the quantum properties of the moving electrons are disregarded, there is no energy conservation, neither entanglement between matter and the irradiated field \cite{Aharonov1,Aharonov2,collins2024conservation}. Consequently, in this semi-classical picture, superpositions of Fock states can be created and the particle number SSR is not explicit. It is important to recall that all the states and operations in the CV representation have an equivalent representation in the form \eqref{Psi}. 

Coherent states are a notable example of ``classical-like" states, according to, up to our knowledge, every possible criterium of non-classicality, physical or computational.  However, they are not SSRC states, leading to the issue of defining a classical-like state in this representation. As detailed in the Supplementary Material \cite{SM}, a coherent state $\ket{\alpha}_{\bf A}$ corresponds to the $N \rightarrow \infty$ limit of state $\ket{0}_{\bf A}\ket{N}_{\bf R}$. It is then tempting to define Fock states $\ket{N}$ as classical-like states. To further support this position - which is unusual in the CV picture since Fock states and non-Gaussian -, we notice that \eqref{Psi} can be mapped into a spin-$N/2$ system, and also into a symmetric state of $N$ spin $1/2$-like particles \cite{Schwinger2001, Sakurai:94,Majorana1932AtomiOI} (see also \cite{SM} for a recap). In these systems, a Fock state $\ket{0}_{\bf A}\ket{N}_{\bf R}$ indeed corresponds to a classical-like state (or a spin coherent state \cite{Byrnes_Ilo-Okeke_2021, perelomov_generalized_1977,JMRadcliffe_1971, PhysRevA.6.2211}) for being a separable state. Finally, Fock states have been shown to be resourceless according to different consistent definitions \cite{PhysRevX.10.041012, PhysRevLett.131.030801,Extremal,acoherence}.

The correspondence between the SSRC representation and spin systems can reveal previously unexplored implications, which we now investigate. A key finding is that the tools developed in  \cite{PhysRevLett.132.153601, PhysRevLett.127.010504, PhysRevA.108.022424} for symmetric spin-like and angular momentum systems can now be used to achieve universality and, potentially, quantum advantage in general quantum optical systems. This represents a new perspective compared to the widely adopted paradigm introduced in \cite{Braunstein}. Specifically, as demonstrated in \cite{PhysRevLett.132.153601}, the angular momentum operators $\hat J_{x/y/z}$ and their quadratic powers generate a universal set of gates in ${\cal H}_{\bf A}\otimes {\cal H}_{\bf R}$ within symmetric subspaces of fixed $N$ (that in \cite{PhysRevLett.132.153601} corresponds to the number of spin-$1/2$ particles) \cite{Note2}. For a state like \eqref{Psi}, the angular momentum operators are defined as $\hat J_z = (\hat a_{\bf A}^{\dagger}\hat a_{\bf A}- \hat a_{{\bf R}}^{\dagger}\hat a_{{\bf R}})/2$, $\hat J_x= (\hat a_{\bf A}^{\dagger}\hat a_{{\bf R}}+\hat a_{{\bf R}}^{\dagger}\hat a_{\bf A})/2$, and $\hat J_y=i (\hat a_{\bf A}^{\dagger}\hat a_{{\bf R}}-\hat a_{{\bf R}}^{\dagger}\hat a_{\bf A})/2$. We conclude that with a number of such gates that scales polynomially with $N$, we can implement arbitrary transformations in \eqref{Psi}, {\it i.e.}, transformations connecting arbitrary states in the SSRC representation. In quantum optics, the unitaries involving operators $\hat J_{x/y/z}$ correspond to linear optical elements such as beam-splitters (BSs) and phase shifters, which are Gaussian operations \cite{UnitaryMode, Scott}. On the other hand, those involving the quadratic terms $\hat J_{x/y/z}^2$ (and higher powers) correspond to non-linear, Kerr-like interactions, which are non-Gaussian gates and are essential for achieving universality and quantum advantage in the CV framework \cite{Braunstein,PhysRevA.82.052341}. The set of universal gates in the SSRC framework is a subset of the universal gates in the CV framework, which additionally includes squeezing and displacement operators - these, however, do not conserve photon number and are not SSRC. Interestingly, despite these differences, both sets enable the description of the same states and physical phenomena, though viewed from distinct perspectives. 

We now identify a consistent criterium of classicality and non-classicality of quantum states by establishing a mapping between a general SSRC state and a BQC. We will do so by employing an extraction protocol inspired by \cite{PlenioExtract}, that permits converting {\it indistinguishable} particle states into {\it distinguishable} modes states. Notably, this protocol transforms particle separability into modal separability  \cite{PhysRevX.10.041012}.  Utilizing the proposed extraction protocol, arbitrary quantum optical states in the SSRC representation can be converted into states of a $N$ qubit BQC. Consequently, by applying the concepts of universality and quantum advantage developed for BQCs \cite{KLM}, it is possible to analyze and classify general quantum optical states in the SSRC representation - and their CV counterparts - based on a mapping into the BQC framework.
 
The extraction protocol proceeds as follows: mode ${\bf A}$ can be expressed in terms of $2N$ orthogonal modes, so $\ket{N}_{\bf A} \equiv \ket{N}_{\vec{q}}=(\hat{a}^{\dagger}_{\vec{q}})^N/\sqrt{N!}\ket{\emptyset}$, where $\vec{q} \in \mathbb{C}^{2N}$ and $\vec{q} = (q_1,\cdots,q_N,\overline{q}_1,\cdots,\overline{q}_N)^T$ such that \cite{TrepsModes}
\be\label{lala}
\hat{a}_{\vec{q}} =\sum_{i=1}^N q_i\hat b_i(0)+\overline q_i\hat b_i(1).
\ee
Operators $\hat b_{i}(0)$, $\hat b_{i}(1)$ are then anihilation operators associated to a mode $i$ (for instance, the propagation direction) and to modes labelled $0$ and $1$ (as polarization, or another spatial mode \cite{DualRail}). We have that $\sum_{i=1}^N |q_i|^2 +|\overline q_i|^2=1$, and $[\hat{a}_{\vec{q}},\hat{a}^{\dagger}_{\vec{w}}]= \sum_{k=1}^{N} q_k^*w_k + \overline{q}^*_k \overline{w}_k$, where $\vec{w}$ is an arbitrary mode that can also be expanded in an analogous $2N$ dimensional orthogonal mode basis \cite{SM}.  

The expansion of mode $\vec{q}$ in a $2N$ dimensional mode basis plays a crucial role in converting particle properties into modal ones \cite{PlenioExtract}, as the number of modes now matches the number of particles multiplied by two possible internal states for states in the SSRC representation \eqref{Psi}. To conclude the conversion protocol, we define the projector $\mathcal{P}_{\mathcal{S}_N} = \prod_{k=1}^{N} \mathcal{P}_k$, where $\mathcal{P}_k  = \hat{b}_k^{\dagger}(0)\ket{\emptyset}\bra{\emptyset} \hat{b}_k(0) + \hat{b}_k^{\dagger}(1)\ket{\emptyset}\bra{\emptyset} \hat{b}_k(1)$, and ${\cal S}_N$ labels the subspace generated by the projection $\mathcal{P}_{\mathcal{S}_N}$. We have,
\begin{eqnarray} \label{comp}
&&\mathcal{P}_{\mathcal{S}_N} [\hat{a}^{\dagger}_{\vec{q}}]^N\ket{\emptyset} = 
\prod_{k=1}^{N} \left[q_k^* \hat{b}_k^{\dagger}(0) + \overline{q}_k^* \hat{b}_k^{\dagger}(1)\right] \ket{\emptyset} \propto \nonumber \\
&&  \prod_{k=1}^N(\cos{\theta_k}\hat b_k^{\dagger}(0)+e^{i\varphi_k}\sin{\theta_k}\hat b_k^{\dagger}(1))\ket{\emptyset}.
\end{eqnarray}
A single mode state $\ket{N}_{\vec{q}}$, describing a separable particle state is, as announced, converted into a mode separable state. Moreover, \eqref{comp} is a separable state of a BQC, where each external mode $k$, $(k=1,2,\cdots,N)$ labels a qubit prepared in state $(\cos{\theta_k}\hat b_k^{\dagger}(1)+e^{i\varphi_k}\sin{\theta_k}\hat b_k^{\dagger}(1))\ket{\emptyset}$. The $2N$ modes $\{k\}\times \{0,1\}$ are orthogonal, $[\hat{b}_k(p),\hat{b}_{k'}(p')] = 0$ and $[\hat{b}_{k}(p),\hat{b}^{\dagger}_{k'}(p')] = \delta_{kk'}\delta_{p p'}$. With this encoding, the computational basis $\{\ket{x}\}$ is given by
\beq 
\ket{x} = \prod_{k=1}^{N}\hat{b}^\dagger_k(x_k)\ket{\emptyset},
\label{eq:compBasis}
\eeq
where $x\in \{0,\cdots, 2^{N}-1 \}$ and $x_1x_2\cdots x_k \cdots x_{N}$ is the binary representation of $x$. Notice that the above projection transforms particle separability into modal ones \cite{PhysRevX.10.041012} without creating additional resources.

In the conversion protocol described above we have considered a single mode Fock state $\ket{N}_{\vec{q}}$. As we'll see in the following, and in \cite{SM}, the protocol can be executed for arbitrary states in the form $\ket{n}_{\vec{q}}\ket{N-n}_{\vec{w}}$, where now $\vec{w}$ is a mode orthogonal to $\vec{q}$. Mode $\vec{w}$ can be formally identified to the mode ${\bf R}$ in \eqref{Psi}, that serves as a phase reference. Finally, the described mapping can also be applied to superpositions $\ket{\Psi}=\sum_{n=0}^{n} c_n\ket{n}_{\vec{q}}\ket{N-n}_{\vec{w}}$, {\it i.e.}, to states as \eqref{Psi}. In this case, the coefficients $c_n$ are directly related to the non-classical properties of states in the SSRC representation and of the BQC, as entanglement, as we'll see in the following.

As previously discussed, by applying universal gates in the SSRC representation to an initial state $\ket{N}_{\vec{q}}$, we can arbitrarily approach states with any (normalized) complex coefficients $c_n$. We will now see that these gates can also be converted by the extraction protocol into universal gates acting on the BQC. Hence, we propose to use the computational resources created in the BQC to benchmark the non-classicality of the SSRC states, an analysis that can later be extended to the CV representation. 

We now systematically study the effect of universal SSRC gates on an initial state $\ket{N}_{\vec{q}_1}=(\hat a_{\vec{q}_1}^{\dagger})^N/\sqrt{N!}\ket{\emptyset}$, with $\hat a_{\vec{q}_1}^{\dagger}=\sum_{i=1}^N \hat b_{i}^{\dagger}(0)/{\sqrt{N}}$. This choice is of course arbitrary, but serves as a working example. Using the conversion protocol, the BQC is  initialized in the corresponding separable state $\ket{x=0}=\prod_{i=1}^N \hat b_{i}^{\dagger}(0)\ket{\emptyset} \propto \hat {\cal P}_{{\cal S}_N}\left (\hat a_{\vec{q}_1}^{\dagger}\right )^N \ket{\emptyset}$.

We start by applying SSRC Gaussian passive operations, and analyzing their effects both on the SSRC state and in the BQC using the conversion protocol. These gates can be written in terms of rotations generated by the angular momentum operators $\hat J_{x/y/z}$, as previously introduced, by identifying ${\bf A} \rightarrow \vec{q}_1$ and ${\bf R} \rightarrow \vec{w}$, where $\vec{w}$ is orthogonal to $\vec{q}_1$. This means that we consider here for simplicity the cases where the BS-like operators couple two orthogonal modes, but the more general situation where gates couple two arbitrary modes with a non-zero overlap does not present a conceptual difference. 

We can study for example the rotation $e^{2i\eta \hat J_x}\left ( \hat a_{\vec{q}_1}^{\dagger}\right ) ^N \ket{\emptyset} = \left (\cos \eta \hat a_{\vec{q}_1}^{\dagger} + \sin{\eta}\hat a_{\vec{w}}^{\dagger}\right )^N\ket{\emptyset} = \left ( \hat a_{\vec{q}}^{\dagger}\right ) ^N\ket{\emptyset}$ (see Eq.\eqref{lala}). A proper choice of $\eta$ and of the mode $\vec{w}$ permits the transformation of the initial mode $\vec{q}_1$ into any other mode $\vec{q}$. Rotations around arbitrary directions are mode manipulations (in the example, $\vec{q}_1 \rightarrow \vec{q}$) that preserve the quantum state (in the example, a Fock state $\ket{N}$) and its {\it intrinsic} modal properties (in the example, a single mode state remains single mode). This is an alternative way to state that using rotations alone we cannot reach universality in SSRC framework, since they do not permit approaching arbitrary amplitudes $c_n$. This fact can be interpreted in computational terms in the BQC space by using \eqref{comp}: it shows that different modes $\vec{q}$ can be associated to a different {\it product} state of the BQC. Hence, rotations exploring the mode space $\vec{q}$ correspond to {\it local} operations on the BQC that preserve separability and cannot access its full quantum computational power \cite{KLM, SM}.

We can now unify previously disconnected perspectives on the lack of quantum computational advantage. Rotations generated by angular momentum-like operators preserve the modal structure, as they merely represent a change in the mode basis while maintaining the same initial state from the SSRC perspective. Furthermore, these rotations correspond to local operations within a BQC. Specifically, starting with the state $\ket{N}$, restricting to rotations preserves the state's intrinsically single-mode structure, maintaining the separability of the BQC. In \cite{SM}, we provide an example of a local manipulation of two qubits and show the reciprocal: any local operation in the BQC can be represented as a mode transformation $\vec{q}_1 \rightarrow \vec{q}$.

We now move to non-Gaussian (nG) operations, as for instance $e^{4i \eta  \hat J_{z}^2 }=e^{i\eta( \hat n_{\vec{q}_1}-\hat n_{\vec{w}})^2 }$, that can be applied  to a rotated state $\ket{N}_{\vec{q}}$, where  $\hat a_{\vec{q}}^{\dagger}= u_1\hat a_{\vec{q}_1}^{\dagger}+u_w\hat a_{\vec{w}}^{\dagger}$, $|u_1|^2+|u_{\vec{w}}|^2=1$. We have 
\begin{eqnarray}\label{J2}
&&e^{4i\eta  \hat J_{z}^2 }\ket{N}_{\vec{q}} \propto e^{4i\eta  \hat J_{z}^2 }\left (u_1\hat a_{\vec{q}_1}^{\dagger}+u_w\hat a_{\vec{w}}^{\dagger}\right )^N\ket{\emptyset}=  \\
&&\sum_{n=0}^N\binom{N}{n}\left ( u_1e^{i\eta n}\hat a_{\vec{q}_1}^{\dagger}\right )^n \left (u_w e^{i\eta(N-3n)} \hat a_{\vec{w}}^{\dagger}\right )^{N-n}\ket{\emptyset}.\nonumber
\end{eqnarray} 
State \eqref{J2} cannot in general be transformed by rotations into a single mode state. It is a nG mode entangled state that is analogous to a spin-squeezed state \cite{PhysRevA.47.5138}. 

To better analyze how the unitary $e^{4i\eta  \hat J_{z}^2 }$ can create mode entanglement and highlights the importance of a phase reference, we study the example $N=2$ (the case of arbitrary $N$ is studied in \cite{SM}, and follows the same principles as for $N=2$). We use, for instance, $\hat a_{\vec{q}}=1/\sqrt{2}\left ( \hat a_{\vec{q}_1} +\hat a_{\vec{w}} \right )$ (which is a $\hat J_x$ eigenstate), leading to $e^{4i\eta  \hat J_{z}^2 }\ket{2}_{\vec{q}}=1/2\left ( e^{i4\eta}\ket{2}_{\vec{q}_1}\ket{0}_{\vec{w}}+e^{i4\eta}\ket{0}_{\vec{q}_1}\ket{2}_{\vec{w}}+\sqrt{2} \ket{1}_{\vec{q}_1}\ket{1}_{\vec{w}} \right )$. This state cannot be written as a single mode state unless $\eta = n\pi/4$, $n \in \N$.  Physically, the quadratic interaction \eqref{J2} corresponds to a cross-Kerr one \cite{Review1, Review2, NLBS, Scala:2023sti}, see also \cite{Note}.

The non-classical properties of this final state and, more generally, states as \eqref{J2}, can be extracted and converted into quantum computational resources of the BQC, which provides an alternative way to assess them. Using, for instance, in the previous example, $\hat a_{\vec{q}_1}=1/2\left (  \hat b_1(0)+ \hat b_2(0)+  \hat b_1(1)+ \hat b_2(1) \right )$ and $\hat a_{\vec{w}}=1/2\left (  \hat b_1(0)+ \hat b_2(0)-\hat b_1(1)- \hat b_2(1) \right )$ \cite{Note3}, we find ${\cal P}_{{\cal S}_2}e^{4i\eta  J_{z}^2 }\ket{2}_{\vec{q}} \propto \left ( \hat b_1^{\dagger}(0)\hat b_2^{\dagger}(0)(e^{4\eta i}+1)+ \hat b_1^{\dagger}(1)\hat b_2^{\dagger}(1)(e^{4\eta i}-1) \right )\ket{\emptyset}$. If $\eta = n\pi/4$ the state is separable. Otherwise, it is a particle and mode entangled state of the BQC as well. Moreover, if $\eta =\pi/8$, $e^{4i\eta  J_{z}^2 }$ is mapped into a conditional operation that creates maximally entangled states in the BQC. Combined to rotations, this gate enables the efficient exploration of the full Hilbert space of the BQC by moving it outside of the subspace of separable states, and completes the set of universal gates in the BQC \cite{Review2, KLM, PhysRevA.86.022316, qudit} and \cite{SM}.

We can now summarize and discuss our results. We have shown how using the SSRC representation of quantum optical states followed by an extraction protocol one can establish a sound and consistent criterium of non-classicality in terms of the computational resources created in a BQC: while single mode Fock states and their manipulation by Gaussian gates are considered as classical for restricting the BQC to a separable space - so, a system that can be efficiently classically simulated - non-classical states are a resource involving the application of non-Gaussian mode entangling gates in the BQC. Non-classical states promote the BQC to universality when projected into its subspace, potentially leading to the impossibility of its efficient classical simulation. A consequence of these facts is that, constructing a universal BQC where physical qubits encode logical qubits requires, as a resource, Fock states in different modes and a non-classical state - according to the established criterium. Other logical bosonic encodings are possible, and the systematic correspondence established here also enables their interpretation in terms of BQCs. This can be done by representing the different encodings (as \cite{gottesman_encoding_2001}, for instance) in the SSRC framework, followed by the extraction protocol. This procedure reveals how each encoding corresponds to different BQC states. The example of cat-codes \cite{Ofek2016ExtendingTL}, briefly discussed in the Supplementary Material \cite{SM}, demonstrates how a continuous-variable error-correcting code can be directly translated into the definition of logical qubits formed from a collection of physical qubits of a BQC.

One novel and appealing aspect of the presented approach lies in its ability to unify previously distinct hardwares for quantum information processing - the SSRC one (which can be mapped onto CV), the BQC (which can be mapped onto an abstract quantum computer), and angular momentum systems, that can also be used as qudits. Consequently all these encodings can now share a common reference of what are classical and non-classical physical and computational resources. Our results also bring a new perspective to the roles of Gaussian and non-Gaussian resources in quantum advantage. Gaussian operations primarily involve mode basis transformations, whereas non-Gaussian operations can lead to mode entanglement and promote to universality and to potential quantum advantage in the SSRC representation and in the BQC. Even if classical-like states in this representation are non-Gaussian states, their manipulation by Gaussian gates can be efficiently classically simulated. Using the present picture we naturally retrieve results which are compatible to the ones based on the stellar rank classification of classical simulabilty and quantum advantage \cite{PhysRevLett.130.090602}, but now from a physically motivated picture where the total photon number (the stellar rank for single mode states in the SSRC representation) is directly associated to the maximal number of qubits of a quantum computer (or the minimal encoding).

In conclusion, our results provide essential tools for analyzing general bosonic quantum states in terms of their informational and computational potential, explicitly accounting for the phase reference of quantum states. They strongly suggest that the CV description may not be the most ``convenient fiction" \cite{PhysRevA.55.3195, PhysRevA.58.4244, PhysRevA.58.4247} for studying the potential of symmetric states in quantum information, in particular in quantum optics. Furthermore, our results open the perspectives of employing a unified formalism to investigate the non-classical properties of both quantum optical systems \cite{PhysRevA.100.062129, Bartlett2003RequirementFQ} and atomic/spin systems \cite{Aaron, Extremal, PhysRevResearch.3.033134}, promoting the application of concepts such as coherence, anti-coherence, and spin squeezing \cite{PhysRevA.105.022433, PhysRevA.6.2211, PhysRevA.102.012412, PhysRevA.96.032304, acoherence, PhysRevA.47.5138, PhysRevLett.86.4431, Giraud_2010, PhysRevA.92.031801, Aulbach_2010} in quantum optics and in other areas of quantum information science.

\section*{Acknowledgements}

We acknowledge funding from the Plan France 2030 through the project ANR-22-PETQ-0006.

\onecolumngrid
\appendix

\section{I. SSRC, Dicke states, coherent spin states and the Jordan-Schwinger map}

The SSRC representation allows to identify the set of universal gates that when successively applied can connect two arbitrary states which are relevant for quantum optics. This is done by using angular momentum operations, that can be expressed in terms of two non overlapping modes of the bosonic field. This identification is achieved by noting that the 
$2$-mode ($A,R$) state $\ket{n}_A\ket{N-n}_R$ $(n=0,\cdots,N)$ can be considered as a Dicke state $\ket{J=N/2, m = N/2-n}$ $(m=-N/2,-N/2+1,\cdots,N/2)$, describing the symmetric state of $N$ $2$-state ($A$ and $R$) particles (or, alternatively speaking, $N$ spin $1/2$-like particles prepared in a symmetric state). In state $\ket{J=N/2, m = N/2-n}$, we have that $n$ particles are in one state ($A$) and the $N-n$ other particles are in the other state ($R$). 

We recall that the Dicke state~\cite{PhysRev.93.99}  $\ket{J=N/2, m = N/2-n}$ is the eigenstate of the two collective angular momentum operators $\hat J^2 = \hat J_x^2 + \hat J_y^2 + \hat J_z^2$ and $\hat J_z$:
\begin{align*}
\hat J^2 \ket{J=N/2, m = N/2-n} &= J(J+1)\ket{J=N/2, m = N/2-n} \\
\hat J_z \ket{J=N/2, m = N/2-n} &= m\ket{J=N/2, m = N/2-n}
\end{align*}
where $\hat J_i$ ($i=x,y,z$) is the $i$-th component of the total angular momentum operator associated to a collection of $N$ particles with spin $1/2$: 
$\hat J_i = \frac{1}{2}\sum_{n=1}^N \hat \sigma^{(n)}_i$ $(i=x,y,z)$, where  $\hat \sigma^{(n)}_i$ is the Pauli matrix acting in the Hilbert space of the $n$-th particle. 
The $N+1 = 2J+1$ Dicke states span the Hilbert space of the $N$ 2-state particles states which are symmetric under the exchange of any pair. 

Equivalently, this amounts to consider the Jordan-Schwinger map ~\cite{Jordan_1935,Schwinger2001} defining the angular momentum operators
\begin{align}\label{J}
\hat J_x &=  \frac{1}{2} \left(\hat a_A^\dagger \hat a_R +  \hat a_R^\dagger \hat a_A \right) \nonumber \\
\hat J_y &=  \frac{i}{2} \left(\hat a_A^\dagger \hat a_R -  \hat a_R^\dagger \hat a_A \right) \\
\hat J_z &=  \frac{1}{2} \left(\hat a_A^\dagger \hat a_A -  \hat a_R^\dagger \hat a_R \right) \nonumber
\end{align}
giving $\hat J^2  = \frac{\hat{N}}{2}\left(\frac{\hat{N}}{2} +1\right)$ with
$\hat{N} = \hat a_A^\dagger \hat a_A+ \hat a_R^\dagger \hat a_R$.

The universal set of gates for such  systems has been previously investigated \cite{PhysRevLett.132.153601}. Their infinitesimal generators are given by $\hat J_x, \hat J_z, \hat J_x^2,\hat J_z^2$.
Hence, using the fact that these generators can be expressed as \eqref{J}, it gives us a universal set of gates  which successive applications lead to all unitaries connecting arbitrary states which are relevant for quantum optics and that can be expressed using the SSRC representation.

We also notice that starting from the state $\ket{N}_A\ket{0}_R = (\hat a^\dagger_A)^N\ket{\emptyset}$, which can be seen as the particular Dicke state $\ket{J=N/2,m=-N/2}$, and applying unitary gates generated by $\hat J_z$ and $\hat J_y$ (or $\hat J_x$), we  can obtain all the spin coherent states~\cite{perelomov_generalized_1977} $\ket{\theta,\phi} \propto e^{-i\phi \hat{J}_z} e^{i\theta \hat{J}_y} \ket{N/2,-N/2}$, which are considered as classical states or as  the ``least quantum states" \cite{Extremal, acoherence}. These state are
simply Fock states in a different mode $e^{-i\phi \hat{J}_z} e^{i\theta \hat{J}_y} (a^\dagger_A)^N\ket{\emptyset} \propto  (\hat a^{\dagger}_B)^N\ket{\emptyset}$, where $\hat a_B^\dagger = \sin\left (\frac{\theta}{2}\right ) \hat a_A^\dagger - e^{i\phi} \cos\left ( \frac{\theta}{2}\right )  \hat a_R^{\dagger}$

Hence, in the SSRC representation, single mode Fock states can be considered as classical-like spin states. In addition, they are mapped into Glauber's coherent states in the case where $N \rightarrow \infty$ (see next section), which also corresponds the approximation where the phase reference can be treated as classical.

We notice that in the usual  CV representation, Fock states are not considered as classical states. Indeed they are non-gaussian, with a non-positive Wigner function. It is well known that while non-Gaussianity is a necessary condition to bring quantum advantage, it is not sufficient. The mapping presented in the main text is entirely compatible with these facts.

\section{II. From the the SSRC representation to coherent states}

In this section we consider the number state $\ket{N}$ in order to extract a coherent state structure in the limit of $N\to\infty$. First, we introduce an annihilation operator $\hat b$ such that $\ket{N}_{\bf R}=\frac{(\hat b^\dagger)^N}{\sqrt{N!}}\ket{\emptyset}$. We also introduce a secondary mode with annihilation operator $\hat a$. These two modes are supposed to be orthogonal such that $[\hat a,\hat b^\dagger]=0$. We fix $\alpha\in\C$. Based on these two modes, we define two new collective $N$-dependent modes
\begin{align} \label{nossa}
    \hat b_{(N)}^\dagger=\frac{-\alpha}{\sqrt{N}}\hat a^\dagger +\sqrt{1-\frac{\abs{\alpha}^2}{N}}\hat b^\dagger && \hat a_{(N)}^\dagger=\sqrt{1-\frac{\abs{\alpha}^2}{N}}\hat a^\dagger+\frac{\alpha^*}{\sqrt{N}}\hat b^\dagger
\end{align}
A quick computation allows to verify that $[\hat a_{(N)},\hat b_{(N)}^\dagger]=0$, $[\hat a_{(N)},\hat a_{(N)}^\dagger]=[\hat b_{(N)},\hat b_{(N)}^\dagger]$ and to find the inverse relations
\begin{align*}
    \hat a^\dagger=\frac{-\alpha^*}{\sqrt{N}}\hat b_{(N)}^\dagger+\sqrt{1-\frac{\abs{\alpha}^2}{N}}\hat a_{(N)}^\dagger && \hat b^\dagger =\sqrt{1-\frac{\abs{\alpha}^2}{N}}\hat b_{(N)}^\dagger + \frac{\alpha}{\sqrt{N}}\hat a_{(N)}^\dagger
\end{align*}

We can now compute in the limit of $N\to\infty$
\begin{align*}
    \ket{N}_{\bf R}&=\frac{(\hat b^\dagger)^N}{\sqrt{N!}}\ket{\emptyset}\\
    &=\frac{1}{\sqrt{N!}}\left(\frac{\alpha}{\sqrt{N}}\hat a_{(N)}^\dagger+\sqrt{1-\frac{\abs{\alpha}^2}{N}}\hat b_{(N)}^\dagger\right)^N\ket{\emptyset}\\
    &=\frac{1}{\sqrt{N!}}\sum_{k=0}^N \binom{N}{k}\frac{\alpha^k}{\sqrt{N}^k}\sqrt{1-\frac{\abs{\alpha}^2}{N}}^{N-k} (a_{(N)}^\dagger)^k(\hat b_{(N)}^\dagger)^{N-k}\ket{\emptyset}\\
    &=\sum_{k=0}^N \binom{n}{k}\frac{\sqrt{k!}\sqrt{(N-k)!}}{\sqrt{N!}}\frac{\alpha^k}{\sqrt{N}^k}\sqrt{1-\frac{\abs{\alpha}^2}{N}}^{N-k} \ket{k}_{a_N}\ket{N-k}_{b_N}\\
    &= \sum_{k=0}^N \sqrt{\frac{N(N-1)\cdots(N-k+1)}{k!}}\frac{\alpha^k}{\sqrt{N}^k}\sqrt{1-\frac{\abs{\alpha}^2}{N}}^{N}\sqrt{1-\frac{\abs{\alpha}^2}{N}}^{-k}\ket{k}_{a_N}\ket{N-k}_{b_N}\\
    &\simeq \sum_{k=0}^N \frac{\sqrt{N}^k}{\sqrt{k!}}\frac{\alpha^k}{\sqrt{N}^k} e^{-\abs{\alpha}^2/2}\ket{k}_{a_N}\ket{N-k}_{b_N}\\
    &=e^{-\abs{\alpha}^2/2}\sum_{k=0}^N \frac{\alpha^k}{\sqrt{k!}}\ket{k}_{a_N}\ket{N-k}_{b_N},
    \end{align*}
where we have introduced the notation $(\hat a_{(N)}^\dagger)^k(\hat b_{(N)}^\dagger)^l\ket{\emptyset}=\sqrt{k!l!}\ket{k}_{a_N}\ket{l}_{b_N}$. We see that we can identify mode $a_N$ to {\bf A} and mode $b_N$ to {\bf R} in the main text (see also \eqref{nossa} for $N \rightarrow \infty$). We made the approximations at the limit where $\frac{|\alpha|^2}{N}\ll 1$
\begin{align}
    N(N-1)\cdots(N-k+1)\simeq N^k &&\sqrt{1-\frac{\abs{\alpha}^2}{N}}^N\simeq e^{-\abs{\alpha}^2/2} &&\sqrt{1-\frac{\abs{\alpha}^2}{N}}^k\simeq 1.
\end{align}
Applying the entangling gate with action $\ket{k}_{a_N}\ket{l}_{b_N}\mapsto \ket{k}_{a_N}\ket{k+l}_{b_N}$ we now can identify mode $b_N$ to {\bf K} and $a_N$ to {\bf G} in the main text. We obtain in the limit of large $N$
\begin{equation}
    \ket{N}_{\bf R}\mapsto \left(e^{-\abs{\alpha}^2/2}\sum_{k=0}^\infty \frac{\alpha^k}{\sqrt{k!}}\ket{k}_{a_N}\right)\otimes \ket{N}_{b_N}=\ket{\alpha}_{a_N}\otimes\ket{N}_{b_N},
\end{equation}
where $\ket{\alpha}_{a_N}$ is a the coherent state of amplitude $\alpha$ in the mode $a_N$, associated to the anihilation operator $\hat a_{(N)}$.

\section{III. Mode manipulations on a Fock state}

We detail the mode manipulations implemented on state in the SSRC representation of the main text:
\begin{eqnarray*}
\lvert N\rangle_{{\bf A}}&\equiv& \lvert N\rangle_{\vec{q}}=\frac{\left(\hat a_{\vec{q}}^{\dagger}\right)^{N}}{\sqrt{N!}}\lvert\emptyset\rangle=\frac{1}{\sqrt{N!}}\left(\sum_{n=1}^{N}\left(\hat b_i^{\dagger}(0)q_{i}+\hat b_i^{\dagger}(1)\bar{q}_{i}\right)\right)^{N}\lvert\emptyset\rangle=\frac{1}{\sqrt{N!}}\left(\hat b^{\dagger}(0)+\hat b^{\dagger}(1)\right)^{N}\lvert \emptyset \rangle\\
&=&\sum_{n=1}^{N}\frac{\sqrt{N!}}{\sqrt{n!}\sqrt{(N-n)!}}\frac{\left(\hat b^{\dagger}(0)\right)^{n}}{\sqrt{n!}}\frac{\left(\hat b^{\dagger}(1)\right)^{N-n}}{\sqrt{(N-n)!}}\lvert\emptyset\rangle=\sum_{n=1}^{N}\sqrt{\left(\begin{array}{c}
N\\
n
\end{array}\right)}\lvert n\rangle_{b(0)}\lvert N-n\rangle_{b(1)},
\end{eqnarray*}
where $\sum_{i=1}^N  |q_i|^2 + |\overline{q}_i|^2 = 1$, and we have defined an annihilation operator $\hat a_{\vec{q}}$ in vector space of $2N$ orthogonal modes $\vec{q}\in \mathbb{C}^{2N}$, with $\vec{q}=(q_1,\cdots, q_N, \bar{q}_1,\cdots,\bar{q}_N)^T$:
\begin{eqnarray*}
\hat a_q = \sum_{n=1}^{N} \hat b_i(0)q_{i}+\hat b_i(1)\bar{q}_{i}.
\end{eqnarray*}
The operators $\hat b(0),\hat b(1)$ represent the internal modes (e.g., polarization of the photon) labeled with (0) and (1), each corresponding to an 
$N$-dimensional subspace: $\hat b(0)=\sum_{i=1}^N q_i \hat b_i (0)$, $\hat b(1)=\sum_{i=1}^N \bar{q}_i \hat b_i (1)$.  Correspondingly, the operators $\hat b_i(0(1))$ are the annihilation operators associated  with mode $i$ (for instance, the propagation direction), and they are orthogonal to each other. 

\section{IV. A bosonic quantum computer}

The goal of this section is to familiarize the reader with the fact that, in BQC, information encoding is done in modes occupied by single photons, which are indistinguishable particles. Using this specific photonic state, where each mode is occupied by at most one photon, it is possible to obtain quantum advantage and efficiently calculate in a $2^N$ dimensional $N$ qubit space spanned by the $2N$ modes, the number of modes used to encode $N$ qubits that can occupy two possible states. 

A bosonic quantum computer (BQC) can be used to encode $N$ bits of information into qubits, spanning a $2^N$ dimensional orthogonal space. It was shown in \cite{KLM} that a BQC can efficiently encode and process quantum information using a universal gate set that we discuss in the following. In a BQC,  $N$ photons occupy $2N$ modes, and each mode contains at most one photon. We will label these modes using two numbers, one of them $(i=1,...N)$ that will be used to identify a qubit, and the other $k=0,1$, the qubit's state. An example of a $N$ qubit state is then given by $\ket{x=0}=\prod_{i=0}^N \hat b^{\dagger}_i(0)\ket{\emptyset}$, where $i$ labels the qubit number and $0$ its state. 

Since photons are bosons, they cannot be identified to qubits. So, it is important to recall that information encoding is done in modes, but since these modes are occupied by photons prepared on a specific state, where each mode is occupied by at most one photon, the system behaves as a $N$ qubit system. 

Moreover, the gates manipulating this system can be described in terms of mode manipulations where the modes are treated as $2N$ dimensional qudits. This is clear when studying the system using the first quantization formalism. A two photon state, for instance, can be written as $\ket{x=0}=1/\sqrt{2}(\ket{1,0;2,0}+\ket{2,0;1,0})$, which is a symmetric state of two particles occupying each one mode out of four possible modes it can occupy. We have defined $\ket{i,j}$ as the mode state of a photon, where $i$ is the auxiliary mode, as the propagation direction, that labels the qubit, and $j=0,1$ is the polarization mode, for instance, that labels the qubit's state. In the first quantization notation, we see that each photon can occupy one out of four possible modes.These modes correspond, using the introduced notation, to states $\ket{1,0}$, $\ket{2,0}$, $\ket{1,1}$ and $\ket{2,1}$.

Gates in first quantization are mode manipulations, that can be written as $\hat \sigma_{\alpha, i}$, where $\alpha=x,y,z$, and $i$ corresponds to the mode designating the qubit. For instance, $\hat \sigma_{z,1}= \ket{1,1}\bra{1,1}-\ket{1,0}\bra{1,0}$ is a Pauli matrix $\hat \sigma_z$ acting on the first mode (qubit). Since in first quantization we make explicit that the gates are applied to each photon, we have that $(\hat \sigma_{\alpha, i})_j$ designates the application of $\hat \sigma_{\alpha, i}$ into the $j$-th particle. Hence, the local manipulation of qubit $1$ of state $\ket{x=0}$ with $N$ photons ($N$ spatial modes, each with $2$ polarizations) can be expressed as 
\be
e^{i\beta \hat J_{\alpha, 1}}\ket{x=0},
\ee
where $\hat J_{\alpha, 1}=\sum_{j=1}^{N} (\hat \sigma_{\alpha,1})_j$. In the example, we have studied a mode coupling that is relevant for a BQC. However, since each photon can be in $2N$ different modes, different mode couplings involving these modes could also be engineered using the $SU(2N)$ generators, as for instance, those coupling different spatial modes, as  $\ket{1,1}\bra{2,1}-\ket{2,1}\bra{1,1}$. This type of operation would, of course, act on each photon independently. Nevertheless, these type of operations are not useful in the present context. 

We can however engineer more complex mode couplings that are compatible with the chosen mode encoding. For example, the manipulation of two qubits involves four modes among the $2N$, as follows: $\hat J_{\alpha, 1} +\hat J_{\beta, 2}=\sum_{j=1}^{N} (\hat \sigma_{\alpha,1})_j+ (\hat \sigma_{\beta, 2})_j$. 

As we have seen in the main text, such linear angular momentum operations preserve the state separability. Moreover, in order to achieve universality, one must complete the gate set with quadratic operations, as for instance, $(\hat J_{\alpha, 1} +\hat J_{\beta, 2})^2=(\sum_{j=1}^{N} (\hat \sigma_{\alpha,1})_j+ (\hat \sigma_{\beta, 2})_j)^2 = \sum_{i\neq j}^{N} (\hat \sigma_{\alpha,1})_i(\hat \sigma_{\beta, 2})_j+\sum_{j=1}^{N}(\hat \sigma_{\beta, 2})_j^2+\sum_{j=1}^{N}(\hat \sigma_{\alpha, 1})_j^2$. Hence, quadratic gates can turn out to be mode entangling ones, and consequently qubit entangling gates. However, they are not the perfect analog of a CNOT gate, for being non-Gaussian gates, so the analogous of the non-Clifford gate for angular momentum.

 \section{V. Non-Gaussian entangling gate application for arbitrary $N$}
 
 We would like to generalize the example in the main text to a situation where $2$ spatial modes are manipulated in a SSRC state with $N$ modes (and $N$ photons, which corresponds to a $N$ qubit BQC). For this, we can develop Eq. (3) of the main text in the situation where we expand $\hat a_{\vec{w}}= c_{1}\hat b_{1}(-)+ c_{2}\hat b_{2}(-)$ and $\hat a_{ {\vec{q}_1}}= \sum_{i=1}^N d_{i }\hat b_{i}(+)$, where $\hat b_i(\pm) \propto \hat b_i(0)\pm \hat b_i(1)$ so the only terms that will be relevant when projecting into the ${\cal P}_{{\cal S}_N}$ subspace are those where $N-n \leq 2$ :
 \begin{eqnarray}\label{CZ}
 &&\sum_{n=N-2}^{N}\binom{N}{n}\left ( u_1e^{i\eta n}\hat a_{\vec{q}_1}^{\dagger}\right )^n \left (u_w e^{i\eta(N-3n)} \hat a_{\vec{w}}^{\dagger}\right )^{N-n}\ket{\emptyset} \propto u_1^{N}e^{i\eta N^2}\left (  \sum_{i=1}^N d_i\hat b_{i}^{\dagger}(+)\right )^N +   \\
 &&N u_1^{(N-1)}u_w e^{i\eta (N-2)^2}\left (  \sum_{i=1}^N d_i \hat b_{i}^{\dagger}(+)\right )^{N-1}\left ( c_1\hat b_{1}^{\dagger}(-)+ c_2\hat b_{2}^{\dagger}(-)\right)+ \nonumber\\
&&\frac{N(N-1)}{2} u_1^{(N-2)}u_w^2 e^{i\eta (N-4)^2}\left (  \sum_{i=1}^N d_i\hat b_{i}^{\dagger}(+)\right )^{N-2}\left ( c_1\hat b_{1}^{\dagger}(-)+ c_2\hat b_{2}^{\dagger}(-)\right)^2\ket{\emptyset}.\nonumber
\end{eqnarray}
The projection into a BQC with $N$ photons is proportional to  
\begin{eqnarray}\label{proj}
&&u_1^{N}e^{i\eta N^2}\prod_{i=1}^N d_i\hat b_i^{\dagger}(+) + u_1^{(N-1)}u_w e^{i\eta (N-2)^2}\left ( \prod_{i\neq1}^N d_i\hat b_i^{\dagger}(+) c_1\hat b_{1}^{\dagger}(-)+e^{i\eta (N-2)^2}\prod_{i\neq2}^N d_i\hat b_i^{\dagger}(+) c_2\hat b_{2}^{\dagger}(-)\right )\nonumber \\
&&u_1^{(N-2)}u_w^2 e^{i\eta (N-4)^2} \prod_{i >2}^N d_i\hat b_i^{\dagger}(+) c_1\hat b_{1}^{\dagger}(-)c_2\hat b_{2}^{\dagger}(-) \ket{\emptyset}= \\
&&u_1^{N-2} \prod_{i >2}^N d_i\hat b_i^{\dagger}(+) (e^{i\eta N^2}d_1d_2u_1^2 \hat b_1^{\dagger}(+) \hat b_2^{\dagger}(+)+e^{i\eta (N-2)^2}c_1d_2u_1u_w\hat b_{1}^{\dagger}(-) \hat b_2^{\dagger}(+)+\nonumber \\
&&e^{i\eta (N-2)^2}d_1c_2u_1u_w\hat b_{1}^{\dagger}(+) \hat b_2^{\dagger}(-)+e^{i\eta (N-4)^2} c_1c_2u_w^2\hat b_{1}^{\dagger}(-) \hat b_2^{\dagger}(-))\ket{\emptyset}.\nonumber 
\end{eqnarray}
Since the decomposition of $\hat a_{\vec{w}}$ (and $\hat a_{{\vec{q}_1}}$) into orthogonal modes is arbitrary, we can always choose $c_1= c_2=u_1$ and  $d_1=d_2=u_w$ (notice that $u_1$ and $u_w $ are fixed by the gate and the initial mode ${\vec{q}_1}$). In this case, Eq. \eqref{proj} simplifies to something proportional to 
\be\label{simple}
 \prod_{i >2}^N d_i\hat b_i^{\dagger}(+) (e^{i\eta N^2} \hat b_1^{\dagger}(+) \hat b_2^{\dagger}(+)+e^{i\eta (N-2)^2}\hat b_{1}^{\dagger}(-) \hat b_2^{\dagger}(+)+e^{i\eta (N-2)^2}\hat b_{1}^{\dagger}(+) \hat b_2^{\dagger}(-)+e^{i\eta (N-4)^2} \hat b_{1}^{\dagger}(-) \hat b_2^{\dagger}(-))\ket{\emptyset}.\nonumber 
\ee 
We can now perform a change of variables $\widetilde {\hat b}_{1(2)}^{\dagger}(-)= e^{i\frac{\eta (N-4)^2}{2}} \hat b_{1(2)}^{\dagger}(-)$, $\widetilde {\hat b}_{1(2)}^{\dagger}(+)= e^{-i\frac{\eta (N-4)^2}{2}+i\eta (N-2)^2} \hat b_{1(2)}^{\dagger}(+)$. Consequently, the term $\widetilde {\hat b}_1^{\dagger}(+) \widetilde {\hat b}_2^{\dagger}(+)$ has a cumulated phase $\phi = 8\eta$ that does not depend on $N$. Consequently, a gate $e^{i\frac{\pi}{8}\hat J_z}$, with $\hat J_z=(\hat n_{\vec{q}_1}-\hat n_{\vec{w}})/2$, where $\vec{q}_1$ and $\vec{w}$ are arbitrary orthogonal modes with overlap $u_1$ and $u_w $, respectively, with an initial mode ${\vec{q}_1}$ to which the gate is applied, can always be seen as a controlled-Z gate ({\it i.e.}, $\phi=\pi$) between two modes encoding quantum information in one of the Fock states $\ket{N}_{{\vec{q}_1}}$ of the initial mode. The action of this gate is independent of the number of photons $N$, and the modes where information is encoded can always be chosen in a way that illustrates this result. Of course, this gate can be seen from different perspectives by using different mode decomposition, but the goal of this one is simply to illustrate how one can always extract universality and potential quantum advantage for the BQC from such a gate, using a parameter that is always the same ($\eta=\pi/8$) irrespectively of the total number of photons.

\section{VI. Examples of local and non-local manipulations of the BQC}

We provide an illustrative example of how local operations can be performed in a BQC with $N=2$. We define $\hat a_{{\vec{q}_1}} = \frac{1}{\sqrt{2}} \left ( \hat b_{1}(0)+\hat b_{2}(0)\right )$ and $\hat a_{\vec{w}} = \cos{\theta}\hat b_{1}(1)+e^{i\varphi}\sin{\theta}\hat b_2(1)$. For $\theta, \varphi=0$, for instance, we have that $\ket{\psi'}=e^{i\eta \hat J_x}\frac{1}{{2}} \left ( \hat b_{1}^{\dagger}(0)+\hat b_{2}^{\dagger}(0)\right )^2 \ket{\emptyset} = \frac{1}{{2}} \left (\cos{\eta}(\hat b_{1}^{\dagger}(0)+\hat b_{2}^{\dagger}(0))+ \sin{\eta}\hat b_{1}^{\dagger}(1)\right )^2 \ket{\emptyset}$. Hence, ${\cal P}_{{\cal S}_2}\ket{\psi'} = \left ( \cos{\eta}\hat b_{1}^{\dagger}(0)+\sin{\eta}\hat b_{1}^{\dagger}(1)\right ) \hat b_2^{\dagger}(0)\ket{\emptyset}$, which corresponds to a rotation of the qubit encoded in the spatial mode $1$. Setting $\theta, \varphi \neq 0$ leads to the simultaneous and independent manipulation of the polarization associated to each spatial mode: 
\begin{eqnarray}
&&{\cal P}_{{\cal S}_2}\ket{\psi'}= {\cal P}_{{\cal S}_2}\frac{1}{{2}} \left (\cos{\eta}(\hat b_{1}^{\dagger}(0)+\hat b_{2}^{\dagger}(0))+ \sin{\eta}(\cos{\theta}\hat b_{1}^{\dagger}(1)+e^{i\varphi}\sin{\theta}\hat b_2^{\dagger}(1)\right )^2 \ket{\emptyset}= \nonumber \\
&&\left (\cos{\eta}\hat b_{1}^{\dagger}(0)+ \sin{\eta}\cos{\theta}\hat b_{1}^{\dagger}(1) \right ) \left (\cos{\eta}\hat b_{2}^{\dagger}(0)+\sin{\eta}e^{i\varphi}\sin{\theta}\hat b_2^{\dagger}(1)\right )\ket{\emptyset}.
\end{eqnarray}

\section{VII. Implementing arbitrary local operations in the BQC}

We now show that all the local transformations on the BQC can be written as a mode transformation in the SSRC state. A general local rotation of qubits encoded in the spatial mode of single photons can be written as 

\be
\hat G_i = e^{\eta_i \hat b_{i}(0)\hat b_{i}^{\dagger}(1)-\eta_i^*\hat b_{i}^{\dagger}(0)\hat b_{i}(1)+i\zeta_i(\hat b_i^{\dagger}(0)\hat b_i(0)-\hat b_i^{\dagger}(1)\hat b_i(1))},
\ee 
Since $[\hat G_i, \hat {\cal P}_{{\cal S}_N}]=0$, we can either apply $G_i$  to the BQC after the projection $\hat {\cal P}_{{\cal S}_N}$ or before it ({\it i.e.}, it can be described as a manipulation of $\ket{\Psi}$. More generally, we have that an arbitrary local gate $\hat {\cal G}$ can be written as $\hat {\cal G} = \prod_j^N \hat G_j$. Hence, an arbitrary local manipulation of the BQC is given by $\hat {\cal G}\hat {\cal P}_{{\cal S}_N} (\hat a_{\vec{q}_1}^{\dagger})^N\ket{\emptyset}=\hat {\cal P}_{{\cal S}_N} \hat {\cal G}(\hat a_{\vec{q}_1}^{\dagger})^N\ket{\emptyset}=\hat {\cal P}_{{\cal S}_N}(\sum_i^N \lambda_i(\cos{|\eta_i|}\hat b^{\dagger}_i(0)+\frac{\eta_i}{|\eta_i|}\sin{|\eta_i|}\hat b_i^{\dagger}(1)))^N\ket{\emptyset}$, where we have defined $q_i = \lambda_i \cos{\eta_i}$ and $\overline q_i = \lambda_i \sin{\eta_i}$, while $\vec{q}_1$ is the initial mode, as in the main text. Using the notation $\vec{w}_i$ for arbitrary modes orthogonal to ${\vec{q}_1}$, we can write $\hat b^{\dagger}_i(0)=c_{i{\vec{q}_1}}\hat a_{\vec{q}_1}+c_{i \vec{w}}\hat a_{\vec{w}_i}$, leading to $\hat {\cal P}_{{\cal S}_N} \hat {\cal G}(\hat a_{\vec{q}_1}^{\dagger})^N\ket{\emptyset}=\hat {\cal P}_{{\cal S}_N}(\cos (|\eta|) \hat a_{\vec{q}_1}^{\dagger}+\frac{\eta}{|\eta|}\sin(|\eta|)\hat a_{\vec{w}}^{\dagger})^N\ket{\emptyset}= \hat {\cal P}_{{\cal S}_N}e^{i\eta(\hat a_{\vec{w}}^{\dagger}\hat a_{\vec{q}_1}+\hat a_{\vec{q}_1}^{\dagger} \hat a_{\vec{w}})}(\hat a_{\vec{q}_1}^{\dagger})^N\ket{\emptyset}=\hat {\cal P}_{{\cal S}_N}(\hat a_{\vec{q}}^{\dagger})^N\ket{\emptyset}= \hat {\cal P}_{{\cal S}_N}\ket{N}_{\vec{q}}$, which is, as we have seen, an intrinsically single mode state, and $\vec{w}=\sum_{i=1}^N k_i \vec{w}_i$ is a mode with no overlap with $\vec{q}_1$ .

 \section{VIII. The interpretation of encoding qubits in continuous variables from the SSRC representation/BQC perspective}

The mapping between quantum optical states and a BQC considered the simplest possible encoding of information, {\it i.e.}, the one where logical qubits of the BQC correspond to physical qubits. Nevertheless, it is known that other encodings are possible, involving the definition of logical qubits consisting of multi-physical qubits. There encodings may display interesting properties, as enabling error correction strategies. 

In CV, it is also possible to encode information in orthogonal states, that form a two dimensional qubit-like subspace. The choice of these states is motivated by different physical or informational constraints. A well known method to encode qubits in bosonic fields consist of using coherent states with different phases (so that they are almost orthogonal) and Schrödinger cat-like states, as follows:
\begin{align}\label{catcode}
&\ket{0} \equiv \ket{\alpha}  & \ket{1} \equiv \ket{-\alpha} & \nonumber \\
&\ket{+} \equiv \frac{1}{\sqrt{{\cal N}_{+}}}(\ket{\alpha}  + \ket{-\alpha}) & \ket{-} \equiv \frac{1}{\sqrt{{\cal N}_{-}}}(\ket{\alpha}  - \ket{-\alpha})&  ,
\end{align}
where ${\cal N}_{\pm}$ are normalization constants. Since the SSRC adopts a quantized phase reference perspective, general quantum states can be described as Eq. (1) of the main text, 	and $N$ can be finite or $N \rightarrow \infty$. We can see from the previous sections, that in the SSRC representation $\ket{\alpha} \rightarrow \ket{N}_b\ket{0}_a$. We will now study the effect of having a coherent state with a different phase and how coherent states with opposite phases can compare to one another within the SSRC representation. For this, we start by comparing two Fock states in the following modes: 
\begin{align*}
\ket{N}_+=\frac{(\hat b_+^{\dagger})^N}{\sqrt{N!}}\ket{\emptyset}, 
\end{align*}
with
\begin{align}\label{plus}
&& \hat a_+^{\dagger}=\frac{-\alpha^*}{\sqrt{N}}\hat b_{(N)}^\dagger+\sqrt{1-\frac{\abs{\alpha}^2}{N}}\hat a_{(N)}^\dagger && \hat b_+^{\dagger} =\sqrt{1-\frac{\abs{\alpha}^2}{N}}\hat b_{(N)}^\dagger + \frac{\alpha}{\sqrt{N}}\hat a_{(N)}^\dagger
\end{align}
and 
\begin{align*}
\ket{N}_-=\frac{(\hat b_-^{\dagger})^N}{\sqrt{N!}}\ket{\emptyset}, 
\end{align*}
with
\begin{align}\label{minus}
&& \hat a_-^{\dagger}=\frac{\alpha^*}{\sqrt{N}}\hat b_{(N)}^\dagger+\sqrt{1-\frac{\abs{\alpha}^2}{N}}\hat a_{(N)}^\dagger && \hat b_-^{\dagger} =\sqrt{1-\frac{\abs{\alpha}^2}{N}}\hat b_{(N)}^\dagger - \frac{\alpha}{\sqrt{N}}\hat a_{(N)}^\dagger.
\end{align}
By inverting \eqref{plus}, 
\begin{align}\label{invert}
    \hat a_{(N)}^\dagger=\frac{\alpha^*}{\sqrt{N}}\hat b_+^\dagger +\sqrt{1-\frac{\abs{\alpha}^2}{N}}\hat a_+^\dagger && \hat b_{(N)}^\dagger=\sqrt{1-\frac{\abs{\alpha}^2}{N}}\hat b_+^\dagger-\frac{\alpha}{\sqrt{N}}\hat a_+^\dagger,
\end{align}
we can replace \eqref{invert} in \eqref{minus}, which leads to 
\be\label{scalar}
\ket{N}_-=\frac{1}{\sqrt{N!}}((1-2\frac{|\alpha|^2}{N})\hat b_+^\dagger -2\frac{\alpha}{\sqrt{N}}\sqrt{1-\frac{\abs{\alpha}^2}{N}}\hat a_+^\dagger)^N\ket{\emptyset}.
\ee
We notice that, indeed, ${}_+\langle N|N\rangle_- = (1-2\frac{|\alpha|^2}{N})^N \simeq e^{-2|\alpha|^2}$ in the limit of $|\alpha|^2 \ll N$, as expected for two coherent states. This is compatible with the calculations in section II, since when $N \rightarrow \infty$ both $\ket{N}_{\pm}$ tend to coherent states with opposite phases. Still in this limit, and in the approximation of Fock states by coherent states, the scalar product \eqref{scalar} goes to $0$ for large $|\alpha|^2$ (we must always have $|\alpha|^2<N$), and this is a known result for coherent states in the CV representation. This means that, in this limit, the modes $\hat b_+$ and $\hat b_-$ are close to orthogonal. However, we can see from the present calculation that in the limit $|\alpha|^2 \rightarrow \infty$ (which is, in principle, the assumption in the CV description) it is inaccurate to represent \eqref{scalar} by an exponential. As a a matter of fact, the present description sets a limit into how large $|\alpha|^2$ can be with respect to the number of photons of the phase reference, $N$. It also shows why in the CV representation, perfect orthogonality is not possible. Nevertheless,  the present results show that achieving mode orthogonality does not require considering the overlap ${}_+\langle N|N\rangle_- $ exponentially decreasing ({\it i.e.}, using the CV description). As a matter of fact, mode orthogonality can always be achieved by setting $|\alpha|^2=N/2$, which is valid, in particular, in the limit where both $N$ and $|\alpha|^2$ are arbitrarily large (notice that $|\alpha|^2 > N/2$ corresponds to the situation where modes $\hat b_+$ and $\hat b_-$ are exchanged and a global physical meaningless global phase appears in mode $\hat b_+$, so this regime does not change the present discussion). 

The present description does not change the usual physical approach to encoding quantum information into almost orthogonal coherent states, since in such experiments $N$ is always large, and $|\alpha|$ is made sufficiently large, but not comparable to $N$. It provides an enriching perspective of this encoding and shows that the SSRC representation is entirely compatible with it, and shows that quasi-orthogonal coherent states are, in fact, the same state (a Fock state) in quasi-orthogonal modes. By using the SSRC representation, we explicitly see that encoding quantum information in different (close to orthogonal) coherent states actually corresponds to an approximation, using classical-like states, of the encoding information into Fock states occupying different modes. So, in the SSRC representation, encoding quantum information in almost orthogonal coherent states corresponds to making the correspondence $\ket{\alpha} \rightarrow \ket{N}_+$ $\ket{-\alpha} \rightarrow \ket{N}_-$, and $\ket{\pm} \rightarrow 1/\sqrt{2}(\ket{N}_+ \pm  \ket{N}_-)$. This state is equivalent to a NOON state when modes $+$ and $-$ are orthogonal.  

By applying the extraction protocol presented in the main text, we have that $\ket{\pm} \rightarrow 1/\sqrt{2}(\prod_{i=1}^N\hat b_i(0)\ket{\emptyset}\pm \prod_{i=1}^N\hat b_i(1)\ket{\emptyset})$, since $\ket{0}\rightarrow \prod_{i=1}^N\hat b_i(0)\ket{\emptyset}$ and $\ket{1}\rightarrow \prod_{i=1}^N\hat b_i(1)\ket{\emptyset}$. We have then that encoding quantum information in specific states of the field can be interpreted, using the results of the main text, as encoding logical qubits into multiple physical qubits. This opens the perspective of interpreting and comparing properties of different encodings, as for instance the possibility of error correction, using a common framework. We leave this for a future work, together with the detailed investigation of this encoding for different regimes of $|\alpha|$ under the BQC perspective. 

\section{IX. Encoding information in modes $\vec{q}$ and $\vec{w}$ and possible mode decompositions.}

We show that the mode decomposition used in the main text, when discussing Eq. (5), is always possible. We start by noticing that it is always possible to find a mode decomposition of $\vec{q}$ in the form $\vec{q}=\sum_i^{N} \frac{1}{\sqrt{2N}}( \vec{q}_i+\vec{\overline q}_i)$, and the same for $\vec{w} =\sum_i^{N} \frac{1}{\sqrt{2N}}( \vec{w}_i+\vec{\overline w}_i)$, with $|\vec{q}_i|^2=|\vec{\overline q}_i|^2=|\vec{w}_i|^2=|\vec{\overline w}_i|^2=1$. We can now define modes $\vec{p}=(\vec{q}+\vec{w})/\sqrt{2}$, and $\vec{k}=(\vec{q}-\vec{w})/\sqrt{2}$. Hence, $\vec{p}=\sum_i^N \frac{1}{\sqrt{2N}}(\vec{p}_i+\vec{\overline p}_i)$ and $\vec{k}=\sum_i^N \frac{1}{\sqrt{2N}}(\vec{k}_i-\vec{\overline k}_i)$, with $\vec{p_i}=(\vec{p}_i+\vec{w}_i)/\sqrt{2}$ and $\vec{k_i}=(\vec{p}_i-\vec{w}_i)/\sqrt{2}$. We now associate to modes $\vec{p}_i$ the field operators $\hat b_i(0)$ and to modes $\vec{k}_i$ the field operators $\hat b_i(1)$ \cite{TrepsModes}. Hence, $\hat a_{\vec{q}}= \sum_i^{N} \frac{1}{\sqrt{2N}}(\hat b_i(0) + \hat b_i(1))$ and $\hat a_{\vec{w}}= \sum_i^{N} \frac{1}{\sqrt{2N}}(\hat b_i(0) - \hat b_i(1))$. The mode choice in the main text corresponds to the particular case of $N=2$, and the present discussion 
shows that this choice, which was made for discussing Eq. (5) in the main text is indeed always possible.

\bibliography{BibCollectiveModes.bib}

\end{document}